# A retrospective look at Regge poles

Alessandro Bottino[1]
*University of Torino, Italy*
*Accademia delle Scienze di Torino, Italy*

**Abstract**

*The theoretical motivations that led Tullio Regge to investigate the analytical properties of the scattering amplitude of the collision process between two particles in terms of complex energy and complex angular momentum are briefly reviewed and set in the context of the S-matrix theory that was developed in the late Fifties and early Sixties of the last century, in an attempt to unravel the properties of the strong interaction.*

## I - Introduction

It will soon be sixty years since Tullio Regge started working on the analytical properties of the scattering amplitude of the collision process between two particles [1]. This line of research led him, in 1959, to consider the angular momentum as a complex variable, and consequently to derive those singularities of the scattering amplitude that became universally known as *Regge poles* [2]. That analysis was further pursued and extended in Ref.3, where, in particular, the link between non-relativistic potential scattering and the relativistic formulation provided by the Mandelstam representation [4] was made possible.

After such a long span of time, it may be interesting to look back at the scientific path followed by Regge: its genesis and development to investigate the properties of the scattering amplitude through its singularities in the complex plane of the angular momentum. This is what is briefly done in the present note, by setting Regge's work in the wide and rich research activity developed by the scientific community in the '50 and '60 of the last century to unravel the properties of the strong interaction. Here we skip many complicated technicalities, and rather prefer to indulge in some pedagogical and historical aspects.

The present review is organized as follows. In the second and third sections, we recall the role played by the Theory of the S-Matrix in obtaining dispersion relations in strong-interaction processes. In Sect. 4 we outline the approach adopted by Regge to obtain the analytic properties of the elastic scattering amplitude in potential scattering, together with a short sketch on the phenomenological consequences of Regge poles. Conclusions are presented in Sect. 5. The appendix is devoted to a short review of the Sommerfeld-Watson transformation, a crucial ingredient of Regge's approach.

---

[1] email: aless.bottino@gmail.com; http//:www.alessandrobottino.it

The present note is dedicated to Tullio Regge, whose physical insight and remarkable mathematical skill were so much enlightening for those of us who had the privilege of working with him.

## 2. The Scattering Matrix

A standard way of studying the properties of elementary particles and of their interactions consists in analyzing the features of the various processes occurring when a number of particles (constituting the so-called initial state **α**) collide together and produce a set of particles in a generic final state **β**.

Let us denote by S(α,β) the complex function that gives the transition amplitude for the process α ⟶ β, and by P(α,β) the corresponding transition probability, P(α,β)≡|S(α,β)|². Summing over all possible final states and normalizing the total probability to the unity one has the constraint

$$\sum_{\beta} P(\alpha,\beta) \equiv \sum_{\beta} |S(\alpha,\beta)|^2 = 1. \qquad (1)$$

The matrix whose elements are S(α,β) is called the scattering matrix or simply the S-matrix, and Eq.(1), together with the further conditions

$$\sum_{\gamma} S(\alpha,\gamma) S^*(\beta,\gamma) = 0 \quad (if\ \alpha \neq \beta)\ , \qquad (2)$$

represents its unitarity properties.

Any amplitude S(α,β) can be worked out completely when the underlying interaction responsible for the process is described by a definite Hamiltonian which allows for a convergent perturbative expansion of S(α,β) in terms of powers of the relevant coupling constant. This is the case of Quantum Electrodynamics, i.e. the theory which provides a quantum mechanical treatment of the electromagnetic interaction occurring between the photon and the matter spinor fields (leptons and quarks). In QED the perturbative expansion is in terms of powers of the electric charge e, and its convergence is guaranteed by the fact that $e^2/2hc \cong 1/137$ (h and c denote Planck's constant and the speed of light, respectively).

In the late Fifties the treatment of the strong interaction was still in troubled waters. In fact, one realized that if the nucleon-nucleon interaction was described *à la Yukawa*, i.e. by the exchange of a pion, with a pion-nucleon coupling constant g, one obtained numerically $g^2/2hc \cong 15$ , a value inadequate for a perturbative expansion of the transition amplitudes.

Due to these difficulties, for the strong interaction one then resorted to a completely different approach, and developed what became denoted as the Theory of the S-matrix [5].

## 3. The Theory of the S-matrix

The idea was to analyze the S-matrix starting from its fundamental properties, i.e. unitarity, analyticity, crossing symmetry; each one of these being dictated by a specific physical principle.

The property of unitarity was already discussed above, and is provided by Eqs. (1-2).

Analyticity and crossing symmetry deserve special attention, and are hereafter discussed in some details.

### 3.1 Analyticity

The principle of causality states that any event occurring in a physical system at time *t* can only influence the future of the system, with no effect on the behavior of the system for times earlier than *t*.

Most remarkably, the causality principle determines the analytical properties of the S-matrix elements in terms of the energy. This link between causality and analyticity can be seen in a variety of ways. To give an illustration of its physical basis we borrow from Ref. 8 the simple example provided by the diffusion of light by a scatterer.

In case of an incident monochromatic wave propagating along the z-axis, at large distance from the scatterer, the full wave function would be (disregarding the photon spin)

$$\psi \approx e^{i\omega(z/c-t)} + f(\omega,\cos\theta)e^{i\omega(r/c-t)}/r, \qquad (3)$$

where ω is the wave number and *f*(ω,cos θ) is the scattering amplitude. However, to discuss the role of the causality principle, it is convenient to rewrite the previous asymptotic behavior in terms of wave packets (the incident wave plane is replaced by a δ-function):

$$\psi_{wp} \approx \frac{1}{2\pi}\int_{-\infty}^{+\infty} d\omega\, e^{i\omega(z/c-t)} + \frac{1}{r}\int_{-\infty}^{+\infty} d\omega\, f(\omega,\cos\theta)\, e^{i\omega(r/c-t)}. \qquad (4)$$

Now, by restraining to the case of forward scattering, i.e. cos θ = 1 (r = z), and by imposing that the scattered packet does not leave the scatterer before the arrival of the incoming packet, one finds that

$$\frac{1}{\sqrt{2\pi}}\int_{-\infty}^{+\infty} d\omega\, f(\omega)e^{i\omega(z/c-t)} = 0, \quad \text{for} \quad z > ct, \qquad (5)$$

where *f*(ω) ≡ *f*(ω,cos θ=1). Hence the forward scattering amplitude is the Fourier transform of a function $f^{'}(\tau)$ vanishing for τ < 0, i.e.

$$f(\omega) = \frac{1}{\sqrt{2\pi}}\int_{-\infty}^{+\infty} d\tau\, f^{'}(\tau)e^{i\omega\tau} = \frac{1}{\sqrt{2\pi}}\int_{0}^{+\infty} d\tau f^{'}(\tau)e^{i\omega\tau}. \qquad (6)$$

Then it turns out that *f*(ω) can be extended analytically in the upper half of the complex ω-plane.

It is worthwhile to recall that in quantum field theory the analytic properties of the scattering amplitude have their origin in the vanishing of the commutator of two field observables taken at points with a space-like separation [9].

## 3.2 Dispersion relations

The analyticity property of the amplitude $f(\omega)$ allows the derivation of an important relation between the real part and the imaginary part of $f(\omega)$, the so-called dispersion relation, which in turn provides the connection between theory and measurable quantities.

This dispersion relation follows from the application of the Cauchy integral formula to the function $f(\omega)$

$$f(\omega) = \frac{1}{\pi i} P \int_\Gamma d\omega' \frac{f(\omega')}{\omega' - \omega}, \qquad (7)$$

where P denotes the principal value and the contour Γ consists of a semicircle of radius R in the upper half plane of ω and centered in the origin, and a segment (-R,+R) including the value ω along the real axis (the contour Γ is taken counterclockwise). Provided that $f(\omega)$ vanishes at infinity faster than $1/|\omega|$, Eq. (7) may be rewritten as

$$f(\omega) = \frac{1}{\pi i} P \int_{-\infty}^{+\infty} d\omega' \frac{f(\omega')}{\omega' - \omega} \qquad (8)$$

or under the form of a relation between the real and the imaginary parts of $f(\omega)$

$$\mathrm{Re}\, f(\omega) = \frac{1}{\pi} P \int_{-\infty}^{+\infty} d\omega' \frac{\mathrm{Im}\, f(\omega')}{\omega' - \omega}, \qquad (9)$$

what is commonly known as a *dispersion relation*.

In some physical applications it turns out that, for large $|\omega|$, $f(\omega)$ does not have the asymptotic behavior required above. In these cases one can nevertheless derive a dispersion relation by applying the previous procedure to the function $f(\omega)$ divided by a convenient power of the variable ω. For instance, rewriting Eq.(7) for the function $f(\omega)/\omega$ and assuming that $f(-\omega) = f^*(\omega)$ one obtains

$$\mathrm{Re}\, f(\omega) = \frac{2\omega^2}{\pi} P \int_{-\infty}^{+\infty} d\omega' \frac{\mathrm{Im}\, f(\omega')}{\omega'(\omega'^2 - \omega^2)}. \qquad (10)$$

This relation has a glorious antecedent in the dispersion relation obtained independently by Kramers and Kronig [10] in terms of the real and the imaginary parts of the index of refraction [11].

The important role of the dispersion relation (10) emerges when one further considers that $\mathrm{Im}\, f(\omega)$ is connected to the total scattering cross section through the optical theorem, derivable from the unitarity condition (see for instance Ref. 7),

$$\mathrm{Im}\, f(\omega) = \frac{\omega}{4\pi c} \sigma_{tot}(\omega). \qquad (11)$$

Indeed, by inserting Eq.(11) into Eq.(10) one finds

$$\mathrm{Re}\, f(\omega) = \frac{\omega^2}{2\pi^2 c} P \int_{-\infty}^{+\infty} d\omega' \frac{\sigma_{tot}(\omega')}{(\omega'^2 - \omega^2)}. \qquad (12)$$

Then the complex function $f(\omega)$ can be fully reconstructed, once the total scattering cross section is measured.

## 3.3 Crossing symmetry

The postulate of crossing symmetry states that a process containing a particle with four-momentum $p_\mu$ in the initial (final) state is described by a scattering matrix element equal to the scattering matrix element of the process where the given particle is replaced by its antiparticle with four-momentum $-p_\mu$ in the final (initial) state.

Let us, for instance, take a generic process involving 4 particles, represented by the graph of Fig.1 and define the covariant variables $s = (p_a + p_b)^2$, $t = (p_a - p_d)^2$, $u = (p_a - p_c)^2$.

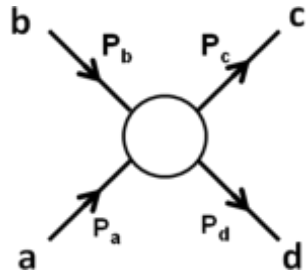

**Fig. 1**

To this graph corresponds a unique S-matrix element which describes the processes:

$$a + b \rightarrow c + d \qquad \text{s - channel}$$
$$a + \bar{d} \rightarrow \bar{b} + c \qquad \text{t - channel}$$
$$a + \bar{c} \rightarrow \bar{b} + d \qquad \text{u - channel}$$

Obviously, these processes involve different regions of the parameter space where the variables *s,t,u* take the physical values for the relevant processes. These invariant variables are linked by the relation

$$s + t + u = m_a^2 + m_b^2 + m_c^2 + m_d^2$$

where the $m_i$'s are the masses of the interacting particles.

## 3.4 Dispersion relations for strong-interaction processes. Mandelstam representation.

The paper by Gell-Mann, Goldberger, and Thirring, mentioned above [9], was crucial in outlining the role of causality in the analytic properties of the S-matrix in quantum field theory. This prompted much research activity in the derivation of dispersion relations in strong-interaction processes. Pioneering works in this directions are those of Ref.12 where dispersion relations for pion-nucleon scattering in the forward direction were first obtained.

Analytic properties of the scattering amplitude in energy at a fixed value of the momentum transfer, and the ensuing dispersion relations, were derived by N. N. Khuri [13] in a non-relativistic approach. The procedure consisted in the application of the Fredholm theory to the scattering integral equation.

A remarkable approach was taken by S. Mandelstam [4], who initiated the investigation of physical processes in more than a single complex variable. In particular, by considering the four-line process of Fig.1 he considered the analytic properties of the generic amplitude as a function of two independent variable, say *s* and *t*. In case of the scattering a + b → c + d, these variables would stand for the squared total center-of-mass energy and the (negative) square of the momentum transverse, respectively. Mandelstam then put forward the conjecture that the two-to-two amplitude can be written as a representation in terms of double integrals, giving also some recipes for the locations of possible singularities. This opened a competition among various scholars to prove the Mandelstam conjecture.

## 4 Analytic properties in potential scattering: Regge's approach

The aim of Regge's line of research, initiated in Ref.2 and continued in Ref. 3, was to check the analytic properties implicit in the Mandelstam's conjecture. Regge's approach was carried out in the context of non-relativistic potential scattering, in the case of a two-body elastic collision. The crucial point of his investigations was the usual partial-wave expansion of the scattering amplitude $f(k,\cos\theta)$ in terms of the phase-shift $\delta(l,k)$ [14-15]

$$f(k,\cos\vartheta) = \frac{1}{2ik}\sum_{l=0}^{\infty}(2l+1)[\exp[2i\delta(l,k)]-1]P_l(\cos\vartheta) \quad (13)$$

that he transformed into the expression

$$f(k,\cos\vartheta) =$$
$$= \frac{1}{2k}\int_{-i\infty}^{i\infty}d\lambda \frac{\exp[2i\delta(\lambda,k)]-1}{\cos(\pi\lambda)}P_{\lambda-\frac{1}{2}}(-\cos\vartheta)\lambda + \frac{i\pi}{2k}\sum_{n=1}^{N}S_n P_{l_n}(-\cos\theta)\frac{2l_n}{\sin(\pi l_n)} \quad (14)$$

by applying to the series of Eq.(13) the Sommerfeld-Watson procedure (reviewed briefly in the Appendix of the present paper). Regge's winning idea of applying the Sommerfeld-Watson procedure to the partial-wave expansion of Eq.(13) led him naturally to the introduction of a complex angular momentum, in exactly the same way as, in Sommerfeld's derivation, the series index n is replaced by the complex variable ν in going from Eq.(A1) to Eq.(A3) of the Appendix. In Eq.(14) $\lambda \equiv l + 1/2$ and $S_n$ stands for the residue of $\exp[2i\delta(\lambda,k) - 1]$ at the pole at $\lambda_n = l_n + 1/2$.

It is worth noticing that, notwithstanding the formal similarities, the use of the Sommerfeld-Watson transformation in Refs.[26-29] and in Regge's procedure served quite different purposes. In the former works the aim was to convert a series of slow convergence into one of rapid convergence for the practical need of a fast numerical calculation, whereas in Regge's case the Sommerfeld-Watson transformation was employed to derive analytical properties for the function originally defined by a series in Eq.(13).

Regge's approach required then a detailed analysis of the analytic properties of the phase-shift $\delta(\lambda,k)$ as a function of the two complex variables (angular momentum $\lambda$ and wave number *k*), necessary to validate the transformation of Eq.(13) into Eq.(14), and subsequently to derive, by use of Eq.(14), the analytic properties of the scattering amplitude in terms of the two variables: *k*, cos $\vartheta$, or, alternatively, *s* and *t* (we recall that the wave number *k* and the centre-of-mass scattering angle $\vartheta$ are related to the energy E and to the Mandelstam variables by the relations: $k^2$ = E = s, -t = 2E(1 − cos $\vartheta$)). The full derivation of these properties is very complicated and required the extraordinary mathematical skill of Tullio Regge.

To summarize the main steps and results of the procedure we follow Ref.3, to which we refer for all relevant details.

Given the partial wave Schrödinger equation written in the centre-of-mass system (natural units are adopted: $\hbar = c = 2\mu = 1$, where $\mu$ is the reduced mass of the two-body system)

$$\psi''(r) + k^2\psi(r) - \frac{\lambda^2 - \frac{1}{4}}{r^2}\psi(r) - V(r)\psi(r) = 0, \qquad (15)$$

it is convenient to define two classes of solutions, identified by the behavior that these solutions have either in the origin or at infinity: a) solution $\varphi(\lambda,k,r)$ that in the origin behaves like $\varphi(\lambda,k,r) \approx r^{\lambda + \frac{1}{2}}$, b) solution $f(\lambda,k,r)$ that at infinity behaves like $f(\lambda,k,r) \approx \exp[-ikr]$. These functions are given explicitly in terms of integral equations in Ref. 3. The potential is assumed to be given by a Yukawian representation [16]

$$V(r) = \int_{m>0}^{\infty} d\mu\, \sigma(\mu) \frac{\exp[-\mu r]}{r}. \qquad (16)$$

An important role is then played by the so-called Jost function $f(\lambda,k)$ defined as the Wronskian of the two previous solutions $f(\lambda,k) \equiv W[f(\lambda,k,r), \varphi(\lambda,k,r)]$, since the exponential of the phase-shift appearing in Eq.(14) can be written as

$$S(\lambda,k) \equiv \exp[2i\delta(\lambda,k)] = \frac{f(\lambda,k)}{f(\lambda,-k)} \exp[i\pi(\lambda - \frac{1}{2})]. \qquad (17)$$

The detailed analysis of the analytic properties of the solutions $\varphi(\lambda,k,r)$ and $f(\lambda,k,r)$ leads to the conclusion the Jost function $f(\lambda,k)$ is analytic in the topological product of the whole k-plane cut along the upper imaginary axis with the half plane Re $\lambda > 0$ and that the poles of the function $S(\lambda,k)$ are located in the regions: Re $k > 0$, Im $\lambda > 0$; Re $k < 0$, Im $\lambda < 0$ [17].

These properties, together with the asymptotic behaviors of $S(\lambda,k)$ (derived by an application of the Wentzel-Kramers-Brillouin method) and of $P_{\lambda - \frac{1}{2}}(-\cos\vartheta)$ for large $\lambda$, prove that for any complex value of $\cos(\vartheta)$ the integral in Eq.(14) is convergent and that, consequently, $f(k,\cos\vartheta)$ is analytic in the $\cos\vartheta$ plane with the exception of the cut $\cos\vartheta$ real $> 1$. This enlarges significantly the region known as the small Lehmann ellipse [18], a domain which was derived from the expression (13) without use of the Sommerfeld-Watson transform.

One important point concerns the behavior of the scattering amplitude for large values of $\cos\vartheta$ (or large $t$). In this regime the integral in Eq.(14) vanishes and $f(s,t)$ is asymptotically given by the sum over the poles. Since at large $t$

$$P_{\lambda - \frac{1}{2}}(-\cos\vartheta) \approx (\cos\vartheta)^{\lambda - \frac{1}{2}} \qquad (18)$$

we obtain

$$f(s,t) \approx C(s)\, t^{\alpha(s)} \qquad \text{(for large } t\text{)}, \qquad (19)$$

where both $C(s)$ and $\alpha(s)$ are complex functions; $\alpha(s)$ is the $l_n(s)$ with the largest real part. This derivation is justified in the case of a finite number of poles. As pointed out in Ref. [19], in the case of occurrence of infinitely many poles, one has to investigate whether the sum over their residues converges at all.

We notice the remarkable property that links the location of the pole of $S(\lambda,k)$ having the rightest position in the $\lambda$ complex plane to the asymptotic behavior of the scattering amplitude at large *t*.

By use of Eq.(14) it is also possible to prove that, for fixed momentum transfer, $f(k,\cos\vartheta)$ is holomorphic in the half plane Im $k > 0$ with the exception of simple poles corresponding to bound states, when k is imaginary.

It then follows that Regge's approach actually proves that in a non-relativistic context $f(s,t)$ satisfies the analytic properties assumed by Mandelstam in his double dispersion relations [4].

Actually, from the link between poles and asymptotic behavior of the scattering amplitude an extremely interesting property emerges if one applies crossing symmetry in such a way that the physical roles of *s* and *t* are exchanged. As a matter of fact, in this instance it turns out that a pole in the *t* channel, with a partial wave function

$$S(l,t) = C(t)/(l - \alpha(t)), \qquad (20)$$

generates the asymptotic behavior

$$f(s,t) \approx C(t)\, s^{\alpha(t)} \qquad \text{(for large } s\text{)} \qquad (21)$$

### 4.1 Phenomenological applications of Regge poles

The paper of Ref.3 was finished by the middle of September 1961 and its preprints sent off to the major scientific institutions and individuals. In the introduction of Ref.3 Regge emphasized the most important points where the previous analysis of Ref.2 had been significantly expanded and improved. These results, also presented by Tullio Regge *in anteprima* in the summer of 1961 at an International School in Hercegnovi (Yugoslavia), raised immediately great interest in the community – with a sudden outburst of an extraordinary number of papers elaborating on the implications of the S-matrix singularities in the complex angular momentum plane: *the "fever" of Regge poles phenomenology exploded* [20]. Seminal papers on this matter were written by Geoffrey F. Chew and Steven C. Frautschi at Berkeley (USA) with Stanley Mandelstam [21] at Birmingham (UK), and by Vladimir N. Gribov and Isaak Ya. Pomeranchuk in Soviet Union [22].

A crucial point in the phenomenology based on Regge poles is that the location of the singularities of $f(s,t)$ (discussed at the end of the previous section) caused by poles in the t channel have a location in the $\lambda$ complex plane that varies in terms of *t*. Interpreting the angular momentum *l* as the intrinsic angular momentum, a Regge singularity establishes a relation, j = α(t=m$^2$), between spin and mass of the exchanged particle.

One then obtains a physical description of the process which looks like a generalization of the old Yukawa conjecture that interpreted the interaction between two nucleons as the exchange of a pion in the transverse momentum.

Regge's theory was extensively used to classify, in homogeneous Regge families, particles and resonances which originally appeared somewhat uncorrelated. This gave for instance origin to *Regge trajectories*, which are lines plotting the function j = α(t) in a diagram of the variable j as a function of t=m$^2$. Refs. 23-25 are among the many books discussing various phenomenological aspects related to Regge poles.

## 5 Conclusions

Regge-poles phenomenology developed with great success for a few years as a branch of particle physics research, in part aside from the original theoretical avenue that analyzed mathematically the S-matrix analytic properties. This phenomenology, although still employed nowadays in some applications, underwent a natural decline, when in the early Seventies *Quantum Chromo Dynamics* made its appearance as the fundamental theory for the strong-interaction force. With the advent of QCD, the strong interaction could be unified with weak and electromagnetic interactions in force of the fundamental principle of gauge invariance.

Independently of the historical fate of Regge-poles phenomenology, Regge's approach to the analysis of potential scattering in terms of complex angular momentum remains as a formidable breakthrough to establish crucial properties of the scattering amplitude, i.e. its analytic properties and asymptotic behaviors. Furthermore, Regge's approach had a quite remarkable role in putting the seeds of further developments: the Veneziano model and the ensuing dual models.

## Appendix
## The Sommerfeld-Watson transformation

This appendix is devoted to a brief summary of a physical-mathematical procedure that was employed by many authors [26-29] to solve the classical problem of transmission of Hertzian waves along Earth's surface. To solve this physical problem one is essentially led to discuss a mathematical quantity given by a series of the form

$$g(x, \cos\vartheta) = \sum_{n=0}^{\infty} (2n+1) P_n(\cos\vartheta) g_n(x). \tag{A1}$$

The main difficulty found by the authors of Refs.[26-29], who wanted to obtain an accurate numerical evaluation of the function g(x,cos θ), consisted in the fact that the convergence of the series in Eq.(A1) is very poor, due to the nature of the function $g_n(x)$ (not given explicitly here). To overcome this difficulty, they applied the trick of replacing *n* by a complex variable ν and by rewriting the series in the form of a complex integral defined along an appropriate contour A. By conveniently modifying the contour A in the complex plane of the variable ν and by applying the theory of residues, one finally obtained an expression of the function g(x,cos θ) as a new series whose fast convergence allowed a very accurate evaluation of g(x,cos θ) in terms of just its first few terms (at variance with the series of Eq.(A1) which required the evaluation of thousands of terms).

Hereby we sketch the main steps of the involved mathematical procedure, following the derivation illustrated by Arnold Sommerfeld in one of his famous textbooks [29], which reflected the renowned lectures he gave in Munich.

The first step consists in promoting the summation index $n$ to become a complex variable $\nu$ and in converting the series of Eq.(A1) into an integral over the contour A to be taken clockwise around the points $\nu$ = 0,1,2,3,… (see Fig.2). The result is

$$g(x,\cos\vartheta) = \frac{i}{2}\int_A d\nu \, \frac{2\nu+1}{\sin(\pi\nu)} P_\nu(-\cos\theta) g_\nu(x), \qquad (A2)$$

as can be proved by applying the residue theorem to the integral along the loop A and by considering that the integrand has poles of the first order when $\nu=n$ with a residue $(-1)^n/\pi$.

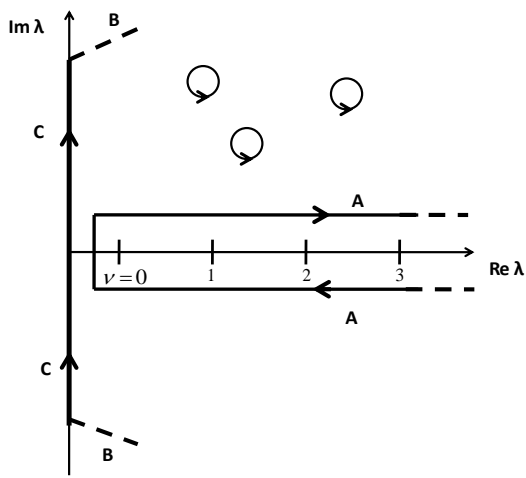

**Fig. 2**

The second step consists in the deformation of the contour A into the vertical path C and the path B connecting paths A and C at infinity.

In the case considered by Sommerfeld the function $g_\nu(x)$ is regular in the $\nu$ right half-plane except for poles in the upper right quadrant. Then the original integral of Eq.(A2) can be rewritten as

$$g(x,\cos\vartheta) =$$
$$= i \int_{-i\infty}^{i\infty} d\lambda \, \frac{\lambda}{\cos(\pi\lambda)} P_{\lambda-\frac{1}{2}}(-\cos\vartheta) g_{\lambda-\frac{1}{2}}(x) - \pi \sum_{\nu_n} \frac{2\nu_n+1}{\sin(\pi\nu_n)} P_{\nu_n}(-\cos\vartheta) \operatorname{Res}[g_\nu(x)]_{\nu=\nu_n} \qquad (A3)$$

once the vanishing of the contribution along the path B is taken into account (as actually proved in Ref. 29). Also, in the integral the variable $\lambda = \nu + \frac{1}{2}$ has been introduced. This change of variable makes the discussion of the integral easier, due to the symmetry property

$$P_{\lambda-\frac{1}{2}}(z) = P_{-\lambda-\frac{1}{2}}(z). \qquad (A4)$$

In the physical case treated in Ref. 29 the integral is proved to vanish; then, the function $g(x,\cos\theta)$, originally given by the series of Eq.(A1), can be calculated in terms of the series of Eq.(A3) whose convergence turns out to be very rapid.